\documentclass[12pt]{article}

\usepackage{graphicx}

\title{ { The Higgs mechanism  from  an extra dimension}}
\author{  Yu.A.Simonov \\
State Research
Center\\Institute of Theoretical and Experimental Physics, \\
Moscow, 117218 Russia}

\newcommand{\beq}{\begin{eqnarray}}
 \newcommand{\eeq}{\end{eqnarray}}
\newcommand{\be}{\begin{equation}}
 \newcommand{\ee}{\end{equation}}

\def\fun#1#2{\lower3.6pt\vbox{\baselineskip0pt\lineskip.9pt
\ialign{$\mathsurround=0pt#1\hfil ##\hfil$\crcr#2\crcr\sim\crcr}}}

\newcommand{{\SD}}{\rm SD}

\newcommand{{\Mc}}{\mathcal{M}}

\begin{document}
\maketitle
\begin{abstract}
The standard $SU(2) \times U(1)$ fields are considered in  4D plus one extra
compact dimension. As a result two basic effects are  obtained. First, four
Goldstone-like scalars are produced, three of them are used to create
longitudinal modes of the $W,Z$ fields, while the fourth becomes the Higgs-like
scalar. Second, $W$ and $Z$ get their masses from the extra compact dimension
with  the standard pattern of symmetry violation. The resulting theory has the
same fields as in the standard model, but without the  Higgs vacuum average.
The properties of the new Higgs scalar and its interaction with fermions are
briefly discussed.
 \end{abstract}

 \section{ Introduction}

The possible role of extra dimensions in the physics of our world was
suggested almost a century ago \cite{1} and is now   a topic of a wide
interest \cite{2,3,4}, see also \cite{5,6} for recent reviews. Possible extensions
of the  Standard Model in general and of the Higgs phenomenon in particular in 5D
are now intensively discussed, often in connection with gravitation \cite{6}
(for a recent study  see  \cite{7}).

 It is a purpose of the present paper to investigate the explicit result of  the
 inclusion  of an extra compact dimension for  the Standard Model Lagrangian
 without Higgs field. In particular, we shall see, that new degrees of freedom
 due to an extra dimension,  $A_{5i}, B_5$, play the same role, as the
 Goldstone-like modes in the Higgs  mechanism, producing longitudinal modes of $W,Z$.
 Moreover, one extra mode can get massive and play the role of the standard
 Higgs boson. All this can happen without introduction of the scalar Higgs
 field, and its vacuum average, and $W$ and $Z$ get their masses due to the
 compact extra dimension as in the Kaluza-Klein (KK) modes.

 In this way one obtains seemingly the same results as in the Standard Model,
 but without the vacuum average of the scalar field. However, some features
 differ, e.g. the interaction of the new scalar field with fermions, which  is
 proportional to fermion masses in the Standard Model, is now universal.

 The plan of the paper is as follows. In next section the case of the $U(1)$
 fields will be  considered, while the section 3 is devoted to the standard
 $SU(2)\times U(1)$ case. In section 4 the inclusion of the scalar mode, its
 properties and interaction with fermions are discussed, The section 5 contains
 conclusions and prospectives.

\section{ The  $U(1)$ Higgs mechanism  from  5D}

We consider below the vector   fields $A_\mu$ which obtain masses due to  an
extra (fifth)  dimension in the case of the local   symmetry $U(1)$, in full
analogy with the  corresponding $U(1)$ Higgs mechanism \cite{8}. We start with
the $D=5$ Lagrangian, containing the gauge fixing  term $\mu, \nu =0,1,2,3,
x_5\equiv y$, \be L_5 =- \frac14 F^2_{\mu\nu} -\frac14 F^2_{5\mu}
-\frac{1}{2\xi} (\partial_\mu A_\mu)^2, \label{1.6}\ee where  \be F_{\mu 5}
=-F_{5\mu} =
\partial_\mu A_5 -\partial_5 A_\mu, ~~ F_{\mu\nu}= \partial_\mu A_\nu
-\partial_\nu A_\mu.\label{2.6}\ee

In what follows wee shall use the 5D $U(1)$  extension of the Lagrangian
(\ref{1.6}), where we introduce  \be A_\nu (x,y) = \exp (i m_5y) A_\nu(x), ~~ A_5 =
i\varphi_5 (x) \exp (im_5 y)\label{3.6}\ee and \be \tilde L_5=- \frac14
|F_{\mu\nu} |^2 -\frac12 |F_{\mu 5} |^2 -\frac{1}{2\xi} |\partial_\mu
A_\mu|^2.\label{4.6}\ee One can see that (\ref{4.6}) acquires the  form \be
\tilde L_5 = \frac12 (\partial_\mu \varphi_5 (x) - m_5 A_\mu(x))^2 - \frac14
F^2_{\mu\nu}(x) - \frac{1}{2\xi} (\partial_\mu A_\mu)^2.\label{5.6}\ee This form
should be compared with the $U(1)$  Higgs Lagrangian \be L_{Higgs} =- \frac14
F^2_{\mu\nu} - \frac{1}{2\xi} (\partial_\mu A_\mu)^2 + |\partial_\mu - ie
A_\mu) \phi|^2- \frac{\lambda^2}{2} (\phi^+
\phi-\frac{\eta^2}{2})^2,\label{6.6}\ee where the  field $\phi$ can be written
in the  standard way \be \phi = \frac{1}{\sqrt{2}} (\eta +\rho+ i
\varphi_H(x))\label{7.6}\ee and the Goldstone field $\varphi_H(x)$ enters in
the combination \be L_{Higgs} = - \frac14 F^2_{\mu\nu} - \frac{1}{2\xi}
(\partial_\mu A_\mu)^2 +\frac12 (\partial_\mu \varphi_H - e\eta A_\mu)^2 +
\Delta \L_{Higgs}, \label{8.6}\ee

$$ \Delta L_{Higgs} =\frac12 (\partial_\mu \rho)^2 - \frac{\lambda^2 \eta^2
\rho^2}{2} + e A_\mu \varphi_H \partial_\mu \rho + \frac12 e^2 A^2_\mu
(\varphi^2_H + \rho^2)+$$ \be + e^2 \eta \rho A^2_\mu - e\rho A_\mu
\partial_\mu \varphi_H - \frac{\lambda^2}{2} \left(\eta \rho (\rho^2 + \varphi^2_H)
+ \frac{(\rho^2 + \varphi^2_H)^2}{4}\right).\label{9.6}\ee

One  can see, that the first three terms in (\ref{8.6}) coincide with $\tilde
L_5$, when we  identify the Goldstone fields $\varphi_5= \varphi_H$, and put
$m_5=e\eta$, while $\Delta L_{Higgs}$ contains extra terms, describing
interaction of the Higgs meson field $\rho$ and the Goldstone field
$\varphi_H$, and $A_\mu$, which are not important in  our further discussion.
One can now calculate the  propagator of the field $A_\mu$, using $\tilde L_5$,
and as in the case of the Higgs Lagrangian (\ref{6.6}) in the zeroth
approximation $e=0$, one obtains in the Landau gauge, $\xi=0$,  \cite{8}
 \be
G_{\mu\nu}^{(0)} = \frac{g_{\mu\nu} - \frac{k_\mu k_k}{k^2}}{k^2}, ~~
G^{(0)}_\varphi = \frac{1}{k^2}. ~\label{10.6}\ee

In the $e^2$ order, writing
\be G_{\mu\nu} = G^{(0)}_{\mu\nu} + G^{(0)}_{\mu\rho}
\prod_{\rho\sigma} G^{(0)}_{\sigma\nu}+...\label{11.6}\ee with $\prod_{\mu\nu}
= e^2 \eta^2 \left( g_{\mu\nu} -\frac{k_\mu k_\nu}{k^2}\right),$  one has
\be G_{\mu\nu} =\frac{ g_{\mu\nu} -\frac{k_\mu k_\nu}{k^2}}{k^2-e^2\eta^2}, ~~
e^2 \eta^2 = m^2_5.\label{12.6}\ee In this way one obtains the Higgs mechanism
in the $U(1)$ case via the fifth dimension dependence, namely, $A_5 (x,y) = i
\varphi (x) \exp (i m_5 y), A_\mu (x,y) = A_\mu (x)\exp (i m_5 y)$, but without the
Higgs meson terms, i.e. the Higgs mechanism without the Higgs vacuum average.

\section{The $SU(2)\times U(1)$ Higgs mechanism  from  an extra dimension}

We now turn to the case of the $SU(2)_L \times U(1)$ symmetry, with the fields
$\hat A_\mu = A_{\mu i} t_i, ~~ t_i =\frac12 \sigma_i$, and $B_\mu$, so that

\be F_{\mu\nu}^A = \partial_\mu\hat  A_\nu- \partial_\nu \hat A_\mu - ig [\hat
A_\mu , \hat A_\nu], \label{13.06}\ee  and  $B_{\mu\nu} = \partial_\mu B_\nu -
\partial_\nu B_\mu$.  It is convenient to unify both fields $\hat A_\mu$ and
$\hat{\frac12} B_\mu$ in one $(2\times 2)$ matrix  $\hat C_\mu \equiv \hat
A_\mu + \hat{\frac12} B_\mu$, where $\hat  1$ is the unit $(2\times 2)$ matrix,
and write \be -\frac12    tr (\hat F^A_{\mu\nu} \hat F_{\mu\nu}^{A^+}) -
\frac12tr \hat  B_{\mu\nu}\hat B^+_{\mu\nu}=-\frac12  tr  \hat C_{\mu\nu} \hat
C^+_{\mu\nu}.\label{13.6}\ee

For the mass generation process we shall specifically need the terms \be
-\frac12 tr \hat C_{\mu 5} \hat C^+_{\mu 5} - \frac12 tr \hat C_{5\mu} \hat
C_{5\mu}^+\equiv \mathcal{L}_m.\label{14.6}\ee

For $\hat C_{\mu 5}$ one can write \be \hat C_{\mu 5} = (\partial_\mu A_{5i} -
\partial_5 A_{\mu i} + g e_{kli} A_{\mu k} A_{5l} ) t_i + \hat{\frac12}
(\partial_\mu B_5 -\partial_5 B_\mu).~ \label{15.6}\ee

As before in  (\ref{3.6}), but now for the $SU(2)$ group, we define for $i=1,2$
\be A_{5i} = i \varphi_i (x) I(y);  A_{53} = i\varphi_3 (x)
I_{B}{(y)},~\label{16.6}\ee

$$ \partial_5 A_{\mu i} (x,y) = i m_A A_{\mu i} (x) I(y) ~~ (i=1,2),$$ \be I(y)
=\exp (im_A y)  = \exp (i\pi n \xi), ~~ \xi =\frac{y}{y_+(m_A)}.~\label{17.6}\ee

Similarly to (\ref{16.6}) one can write \be B_5 =i\varphi_5 (x) I_B (y), ~~
I_B(y) = \exp (im_B y)=\exp ( i\pi n \xi) , ~~ \xi
=\frac{y}{y_+(m_B)}.\label{18.6}\ee

The mixing of $A_{\mu3}$ and $B_\mu$ is defined in the standard way,
introducing fields $Z_\mu$ and electromagnetic $\Gamma_\mu$, \be A_{\mu 3}
=\cos \theta~ Z_\mu+ \sin \theta~ \Gamma_\mu, ~~ B_\mu =\cos \theta ~\Gamma_\mu
-\sin \theta~ Z_\mu.\label{19.6}\ee

We now assume that only $Z_\mu$ has the $y$   dependence, \be \partial_5
A_{\mu3} =i\cos \theta~ m_B Z_\mu I_B; ~~ \partial_5 B_\mu =-i \sin \theta~ m_B
Z_\mu I_B.\label{20.6}\ee As a result, $ \mathcal{L}_m$, Eq. (\ref{14.6}) can
be written as

$$ \mathcal{L}_m  =\frac12 \sum_{i=1,2} (\partial_\mu \varphi_i -m_A A_{\mu
i})^2 + \frac12 (\partial_\mu \varphi_5 + \sin \theta m_B Z_\mu)^2 +$$ \be
+\frac12 (\partial_\mu \varphi_3 - \cos \theta m_B Z_\mu)^2 + \Delta
\mathcal{L}_m \equiv \mathcal{L}_m^{(0)} + \Delta \mathcal{L}_m\label{21.6}\ee
where $ \Delta \mathcal{L}_m$ is  $$\Delta \mathcal{L}_m =\frac12
\sum_{i=1,2,3} g^2 A_{\mu l} \varphi_k e_{lki} A_{\mu n} \varphi_m e_{nmi}+
$$
$$+g \sum_{i=1,2} (\partial_\mu \varphi_i -m_A A_{\mu
i}) A_{\mu l} \varphi_k e_{lki} \frac{I(y)+I^+(y)}{2}+$$  \be +g(\partial_\mu
\varphi_3 - \cos \theta m_B A_{\mu3} ) A_{\mu l} \varphi_k e_{lk3}
\frac{I_B(y)(I^*(y))^2 +I^*_B(y)I^2(y)}{2}.~\label{22.6}\ee

Note, that introducing $I_B(y) =I(y) = \exp( i n \pi \xi)$ and integrating over
$d\xi$ in the cyclic interval $-1\leq\xi \leq 1$, one  has only the first term
in (\ref{22.6}) left nonvanishing,

Now we redefine the Goldstone modes $\varphi_5, \varphi_3$, namely, \be
\varphi_3 =\cos \theta \varphi_Z, ~~ \varphi_5 =-\sin \theta
\varphi_Z\label{23.6}\ee and $ \mathcal{L}_m^{(0)}$ acquires the final form \be
\mathcal{L}_m^{(0)}  =\frac12 \sum_{i=1,2} (\partial_\mu \varphi_i -m_A A_{\mu
i})^2 + \frac12 (\partial_\mu \varphi_Z -m_B Z_\mu)^2 .\label{24.6}\ee

Following the same procedure, as in (\ref{10.6})-(\ref{12.6}), one arrives at
the transverse form of the $W,Z$ propagator,  \be G^{(W)}_{\mu\nu} =
\frac{g_{\mu\nu}- \frac{k_\mu k_\nu}{k^2}}{k^2-m^2_W},~~ G^{(Z)}_{\mu\nu} =
\frac{g_{\mu\nu}- \frac{k_\mu k_\nu}{k^2}}{k^2-m^2_Z},\label{25.6}\ee where \be
m^2_W = m^2_A = m^2_Z \cos^2 \theta,~~ m^2_Z=m^2_B.\label{26.6}\ee In this way
all Goldstone modes $\varphi_1, \varphi_2, \varphi_z$ are eaten  by the now
massive fields $A_{\mu i} , i=1,2$ and $Z_\mu$. The fields $\varphi_i ~(i=1,2)$
and $\varphi_Z$   can be removed by the gauge transformations of the vector fields and
no Higgs-like mesons are left in the Lagrangian.

\section{New scalar mode from  extra dimension}

However, we have lost one mode, since we have replaced $\varphi_3,
\varphi_5$ by one mode  $\varphi_Z$ in (\ref{23.6}). In a more
general case one  can write \be \varphi_3 = \varphi_Z \cos  \theta
+ \varphi_H \sin  \theta , ~~ \varphi_5 =-\sin  \theta \varphi_Z +
\varphi_H \cos\theta.\label{52}\ee Substituting (\ref{52}) into
(\ref{21.6}), one obtains, instead of (\ref{24.6}), the following
result \be \mathcal{L}_m^{(0)'}  =\frac12 \sum_{i=1,2}
(\partial_\mu \varphi_i -m_A A_{\mu i})^2 + \frac12 (\partial_\mu
\varphi_Z -m_B Z_\mu)^2 + \frac12 (\partial_\mu \varphi_H)^2
.\label{53}\ee

Now one can see that again the fields $\varphi_i, i=1,2$ and $\varphi_Z$ can be absorbed by
the fields $A_{\mu i}$ and $Z_\mu$, generating longitudial components, but
$\varphi_H$ appears as a valid physical mode. Neglecting now the  absorbed
modes $\varphi_1, \varphi_2, \varphi_Z$ in $\Delta \mathcal{L}_m,$ one obtains
in (\ref{22.6}) \be \Delta \mathcal{L}'_m =\frac12   g^2 \sin^2 \theta
\varphi^2_H((A_{\mu 1})^2 +  (A_{\mu 2})^2).\label{54}\ee
It is clear that the field $\varphi_H$, having no intrinsic mass term, in
contrast to the Higgs field, acquires the mass through radiative corrections,
e.g. from the quadratically divergent $W$ loops  in (\ref{54}), and the higher order
diagrams, as well as from the interaction of $\varphi_H$ with fermions.



For the following we  need some  modifications of the original expressions
for the field strength and long derivatives, to include the factors coming from
the fifth coordinate. Namely, as in (\ref{17.6}),(\ref{18.6}) we define the 5D
extensions of the fields $A_{\mu i} (x)$ and $B_\mu(x)$.

\be A_{\mu i} (x) \to \hat A_{\mu i} (x,y) = A_{\mu i} (x) I (y), \label{60}\ee
\be B_\mu(x) \to B_\mu (x,y) = B_\mu (x) I_B (y),\label{61}\ee while $A_{5i}
(x,y) $ and $B_5(x,y)$ are given in (\ref{15.6}), (\ref{18.6})

Now the  usual long derivative can be rewritten  in terms of fields
$A_\mu(x,y), B_\mu(x,y)$ as  \be \hat D_\mu =
\partial_\mu - i \hat g_i A_\mu (x,y) t_i - i \hat g' B_\mu(x,y)
\frac{Y}{2},\label{62}\ee where \be \hat g_i = g I^* (y)~ (i=1,2), ~~ \hat g' =
g' I^*_B (y) , ~~ \hat g_3 = I_B (y) (I^*)^2;\label{63}\ee and

\be \hat F^A_{\mu\nu} (x,y) = \partial_\mu \hat A_\nu (x,y) - \partial_\nu \hat
A_\mu (x,y) - i \hat g [\hat A_\mu (x,y) , \hat A_\nu (x,y) ]. ~\label{64}\ee
Note, that expressing $I(y)  = I_B(y) =I(\xi)=\exp (i n \pi\xi)$, all extra
factors in $g, g' $ (\ref{63}) reduce to one factor $I^* (\xi)$.

As a result, the total Lagrangian of vector, scalar and fermion fields can be
written as follows: \be \mathcal{L}= \mathcal{L}_A+\mathcal{L}_B+\mathcal{L}_m
+\mathcal{L}_f,\label{65}\ee \be \mathcal{L}_A+\mathcal{L}_B= -\frac12 tr \hat
F_{\mu\nu}^A \hat F_{\mu\nu}^{A^+} - \frac12 tr \hat B_{\mu\nu}  \hat
B_{\mu\nu}^{+},~ \label{66}\ee

$$ \mathcal{L}_m  =\frac12 \sum_{i=1,2} (\partial_\mu \varphi_i -m_A \hat A_{\mu
i}(x,y))(\partial_\mu \varphi_i^+ - m_A \hat A^+_{\mu i}(x,y))  +$$

 \be
+\frac12 (\partial_\mu \varphi_Z (x,y)-   m_B Z_\mu (x,y))(\partial_\mu
\varphi_Z^* (x,y)-   m_B Z_\mu^* (x,y))+ \frac12 \partial_\mu
\varphi_H\partial_\mu\varphi_H^* +
 \Delta \mathcal{L}'_m,  \label{67}\ee and  $ \Delta \mathcal{L}'_m$ is given in (\ref{54}).
 With the notation
\be \mathcal{L}_A+\mathcal{L}_B= \mathcal{L}_{neutr} + \mathcal{L}_{ch
}+\mathcal{L}_{interf},\label{67a}\ee and using (\ref{19.6}), the  contribution of neutral fields
$B_\mu, A_{\mu3}$ in (\ref{66})  can be written as

 $$\mathcal{L}_{neutr} =-\left\{ \frac12 (\partial_\mu
B_\nu-\partial_\nu B_\mu)^2 + \frac12 (\partial_\mu A_{\nu 3} - \partial_\nu
A_{\mu 3})^2 \right\}= $$ \be =-\left\{ \frac12 (\partial_\mu
\Gamma_\nu-\partial_\nu\Gamma_\mu)^2 + \frac12 (\partial_\mu Z_{\nu } -
\partial_\nu Z_{\mu })^2 \right\}.~\label{68}\ee

For the charged fields and their interference with  neutrals one has \be
\mathcal{L}_{ch} =-\left\{ \frac12 \sum_{i=1,2}(\partial_\mu A_{\nu
i}-\partial_\nu A_{\mu i})^2 +   (gA_{\mu k} A_{\nu l} e_{3kl})^2\right\}
\label{69}\ee and finally $L_{interf}$ is
$$ \mathcal{L}_{interf} =-\left\{ [ \cos \theta (\partial_\mu Z_\nu -
\partial_\nu Z_\mu) + 2 \sin \theta (\partial_\mu \Gamma_\nu-\partial_\nu
\Gamma_\mu) ] g e_{kl3} A_{\mu k} A_{\nu l}+\right.$$\be + \sum_{i=1,2}
(\partial_\mu
  A_{\nu i}-\partial_\nu
A_{\mu i})  e_{kli} g A_{\mu k} A_{li}+ g^2 (\cos \theta Z_\mu+ \sin \theta
\Gamma_\mu) (\cos \theta Z_\nu + \sin \theta \Gamma_\nu)\times\label{70}\ee
$$\left. \times (\delta_{\mu\nu} (A_{\lambda 1} A_{\lambda 1} + A_{\lambda 2}
A_{\lambda 2}) -( A_{\mu 1} A_{\nu 1} + A_{\mu_2 } A_{\nu2} )\right\}.$$

One can see that (\ref{68}),(\ref{69}), (\ref{70}) coincide with the vector
field part of the standard Lagrangian.

In the  fermion part of the Lagrangian, $\mathcal{L}_f$  in
addition to the standard expression, one can write for the quarks \be \mathcal{L}_Q = \bar
Q_L i \gamma_\mu \left(\partial_\mu - ig A_{\mu i} t_i -ig' \frac{\hat 1}{6}
B_\mu \right) Q_L,\label{55}\ee the term, containing the fifth components,
\be \Delta \mathcal{L}_Q =i \bar q_{L,R} \left(\partial_5 - ig A_{53} t_3 -
i\frac{g'}{6} \hat 1 B_5\right) q_{R,L},\label{56}\ee where \be A_{53} = i
(\varphi_Z \cos \theta+ \varphi_H \sin  \theta) I_B (y)\label{57}\ee \be B_5 =i
(\varphi_H \cos \theta - \varphi_Z \sin \theta ) I_B (y).\label{58}\ee

Inserting in (\ref{55}) the terms (\ref{57}), (\ref{58}) for $\mu=5$ and
neglecting $\varphi_Z$, one obtains the following interaction of the field
$\varphi_H$ with $f\bar f$\be\Delta \mathcal{L}_{Hf\bar f} = \bar f_{L,R}
\varphi_H \left(g\sin \theta t_3 +g' \cos \theta \frac{\hat Y_f}{2}\right)
f_{RL},\label{47a}\ee violating  the $SU(2)$ invariance in (\ref{47a}), which,
however, is already violated by different up and down quark masses. In
(\ref{47a}) $g=\bar g \cos \theta, ~ g'=\bar g\sin \theta; ~\bar g =0.74$.

Note the difference between (\ref{47a})  and the standard Higgs -- fermion
interaction $\Delta \mathcal{L}_{stand} = \frac{m_f}{\eta} \bar f f H$, where
$m_f$ is the fermion mass and the Higgs condensate $\eta =246 $ GeV. Thus for the  $(\bar t t)$
quarks the coupling coefficient is 0.2, while in $\Delta \mathcal{L}_{stand}$
it is 0.71. At the same time for lighter quarks our result holds the same,
while the standard coefficient is negligible for all $m_f$ except $m_t$.

 One can see that on the r.h.s. of (\ref{56}) the first term creates the  intrinsic mass
 of the quarks, \be i \partial_5 q (x,y) = m_q q(x, y), \label{59}\ee
and the relation (\ref{59})  automatically violates the  $SU(2)$ symmetry,  if $i\hat \partial_5$
is not  $SU(2)$ doublet. As to the quark mass operator $i\hat \partial_5$, it
can be written in the $SU(2)$ form as $\left(\begin{array}{c} { \cos \phi~ i
\partial_5}\\{\sin \phi~ i \partial_5}
\end{array}\right)$, with  $\phi$
different, being applied to $\left(\begin{array}{c} {\rm up}\\{\rm down}
\end{array}\right)$ or $\left(\begin{array}{c}  {\rm down}\\{\rm up}
\end{array}\right)^+$   states, as it  is done in the Higgs condensate
formalism, however, not lifting the mass difference problem.

Now turning to the fermion term $\mathcal{L}_f$, one  should take into account
that fermions have their own factors $I_f(y)$, which can  differ from the
factors $I(\xi) $ entering the vector fields $W_\mu, Z_\mu$.  As a result  the
corresponding term $\mathcal{L}_f^{(0)}(x,y)$ simply coincides with the
standard 4D  expressions, since the factors $I_f(y)$ cancel in the diagonal
product of fields, and  $I(\xi)$ cancels in the products $\hat g \hat A, \hat g' \hat
B$:

\be \mathcal{L}_f^{(0)} = i \bar f_k \left(\partial_\mu - i\hat g \hat A_{\mu
i} (x,y)  t_i - i\hat g' \hat B_\mu (x,y) \frac{Y_f}{2}\right) \gamma_\mu f_k
(x,y),~~ k=L,R.\label{68a}\ee

However, in the charged fields case, $W_\mu^{(\pm)}$, the fermions and antifermions
may belong to different generations.  This results in different factors $I_k(\xi)$ and $I_{\bar k}(\xi)$ of the
 fermions $f_k$ and $\bar f_k$, and one can see that with  our present definition of
 the norm in the $\xi$ space these matrix elements vanish after integration over
 $d\xi$. Thus in this  approximation
 the mixing between generations does not appear and the CKM matrix is diagonal.
 \section{Conclusion and prospectives}

 We have shown that the extension of the standard higgsless Lagrangian,
 including new components of the vector fields $F_{\mu5}, F_{5\mu}$, as well as
 the fifth coordinate $x_5 =y$ dependence of all fields, $F_{\mu\nu}(x) \to
 F_{\mu\nu} (x,y) ~q(x) \to q(x,y)$, automatically produces Nambu-Goldstone
 modes,  which create longitudinal components of the vector fields,
 while present in $F_{\mu 5}$ derivatives in $y$ produce the masses of $W,Z$
 in the  needed proportion to reconstruct symmetries of the Lagrangian.

 Surprisingly there appears the fourth scalar mode, which acquires the mass
 radiatively via interaction with vectors and fermions. This scalar can contest
 (or complement) the standard Higgs boson. Howeve,r it is different  from the
 latter in the interaction with fermions, which is not proportional to the
 fermion masses.

 The results of the paper open many questions and require
 additional steps to clarify the details of the 5D construction.
 In particular, the aspects of the fermion and (possibly)  vector generations and
 mixings were not touched upon.

 Many aspects of dynamics in the extra dimension were not discussed above, and
 only its circular compactification  was exploited for simplicity, which assumes
 possible KK excitations. This topic is crucial for the quarks and will be treated in
 a separate paper \cite{9}.

 The author is grateful for discussions to A.M.Badalian.

 This work was supported by the grant RFBR 1402-00395.

\end{document}